\documentclass[12pt]{article}\usepackage{amsmath,amsfonts}
 \begin{document}

 \def\const{\hbox{\rm const}}
 \def\diag{\hbox{\rm diag}}
 \def\mod{\hbox{\rm mod}}
 \def\rank{\mathop{\hbox{\rm rank}}}
 \def\Image{\mathop{\hbox{\rm Image}}}
 \def\mes{\mathop{\hbox{\rm mes}}}
 \def\grad{\mathop{\hbox{\rm grad}}}
 \def\ind{\mathop{\hbox{\rm ind}}}
 \def\exp{\mathop{\hbox{\rm exp}}}
 \def\d{\mathop{\hbox{\rm d}}}
 \def\k {\mathop{\hbox{\rm k}}}
\def\Diff{\mathop{\hbox{\rm Diff}}}
 \def\ad{\hbox{ad}}
 \def\Ker{\mathop{\hbox{\rm Ker}}}
 \def\Im{\mathop{\hbox{\rm Im}}}
 \def\id{\mathop{\hbox{\rm id}}}
 \def\Tr{\mathop{\hbox{\rm Tr}}}
 \def\so{\mathop{\hbox{\rm so}}}
 \def\Re{\mathop{\hbox{\rm Re}}}
 \def\Im{\mathop{\hbox{\rm Im}}}
 \def\scirc{\mathbin{\hbox{\scriptsize$\circ$}}}
 \def\M{\mathop{${\mathbb R}^2_1$}}
 \def\sgn{\mathop{\hbox{\rm sgn}}}
 \def\arctanh{\mathop{\hbox{\rm arctanh}}}
\def\bX{\mathop{\hbox{\bf X}}}
\def\bV{\mathop{\hbox{\bf V}}}
\def\bK{\mathop{\hbox{\bf K}}}
\renewcommand{\vec}{\boldsymbol}
\renewcommand{\baselinestretch}{1.7} 
\newtheorem{con}{Conjecture}[section]
\newtheorem{prob}{Problem}[section]
 \newtheorem{prop}{Proposition}[section]
 \newtheorem{coro}{Corollary}[section]
 \newtheorem{lem}{Lemma}[section]
 \newtheorem{theo}{Theorem}[section]
 \newtheorem{defi}{Definition}[section]
 \renewcommand{\theequation}{\thesection.\arabic{equation}}
 \newtheorem{cri}{Sufficient condition}
 \renewcommand{\thecri}{\arabic{cri}.}
 \newenvironment{criterion}{\begin{cri}\rm}{\end{cri}}
 \newtheorem{rem}{Remark}[section]
 \newtheorem{exa}{Example}[section]
 \newtheorem{hyp}{Hypothesis}[section]
 \newenvironment{hypothesis}{\begin{hyp}\rm}{\end{hyp}}
 \renewcommand{\therem}{\thesection.\arabic{rem}}
 \newenvironment{remark}{\begin{rem}\rm}{\end{rem}}
 \newfont{\gothic}{eufm10 at 12pt}
 \title{A class of superintegrable systems of Calogero type}
 \author{Roman G. Smirnov\footnote{Electronic mail: smirnov@mathstat.dal.ca}  \\
   Department of Mathematics and Statistics, 
Dalhousie University \\ Halifax, Nova Scotia,  Canada B3H 3J5 \\
Pavel Winternitz\footnote{Electronic mail: wintern@CRM.UMontreal.CA} \\
Centre de recherches math\'{e}matiques et \\ D\'{e}partement de math\'{e}matiques et statistique \\
Universit\'{e} de Montr\'{e}al, C. P. 6128 - Centre ville \\
Montr\'{e}al, QC, Canada H3C 3J7}
    \date{ }
 \maketitle
\noindent Running title: {\large Calogero type superintegrable systems}

\newpage

\begin{abstract}
 We show that the three body Calogero model with inverse square potentials can be interpreted as a maximally superintegrable
 and multiseparable system in Euclidean three-space. As such it is a special case of a family of systems involving one arbitrary
 function of one variable. 
\end{abstract}

\bigskip 

\noindent Indexing Codes: 02.30Ik, 02.40.Ky, 02.20.Hj

\newpage

\section{Introduction}

 The purpose of this article is to investigate the relation between the rational 
three-body Calogero model in one-dimension \cite{FC69} and superintegrable systems in two and three dimensions 
\cite{NWE90, FMSW65, MSVW67}.

The original (quantum) Calogero model was written in the form 
\begin{equation}
\label{E1}
\begin{array}{l}
 \displaystyle \left\{-\left(\frac{\partial^2}{\partial x_1^2} + \frac{\partial^2}{\partial x_2^2} 
+ \frac{\partial^2}{\partial x_3^2}\right)
+\frac{1}{8}\omega^2[(x_1-x_2)^2 + (x_2-x_3)^2 + (x_3-x_1)^2] \right.
\\ [0.5cm]
\displaystyle \left.+\frac{g_1}{(x_2-x_3)^2} 
 + \frac{g_2}{(x_1-x_3)^2} + \frac{g_3}{(x_1-x_2)^2}\right\}\Psi = E\Psi. 
\end{array}
\end{equation}

Upon introducing the centre-of-mass coordinate $R$ and the Jacobi relative coordinates $\rho$ and $\lambda$ \cite{Ja43}
\begin{equation}
\label{JC}
R = \frac{1}{3}(x_1 + x_2 + x_3), \quad \rho = \frac{1}{\sqrt{2}}(x_1-x_2), \quad \lambda = \frac{1}{\sqrt{6}}(x_1-x_2 - 2x_3)
\end{equation} 
equation (\ref{E1}) was rewritten \cite{FC69} as follows
\begin{equation}
\label{E2}
\begin{array}{l}
\displaystyle \left\{-\left(\frac{\partial^2}{\partial \rho^2} + \frac{\partial^2}{\partial \lambda^2}\right) + \frac{3}{8}\omega^2(\rho^2 + \lambda^2) + \right.\\[0.5cm] 
\displaystyle \left. \frac{1}{2}\frac{g_1}{(\sqrt{3}\lambda - \rho)^2} + \frac{1}{2}\frac{g_2}{(\sqrt{3}\lambda + \rho)^2} + \frac{1}{2}\frac{g_3}{\rho^2}\right\}\Psi = E\Psi,
\end{array} 
\end{equation}
where the motion of the centre-of-mass has been factored out.

A superintegrable system is one that admits more integrals of motion than it has degrees of freedom.
 Systematic searches for superintegrable systems of the form
\begin{equation}
\label{E3} 
H(\vec{x},\vec{p}) = \frac{1}{2}\vec{p}^2 + V(\vec{x})
\end{equation}
have been conducted in Euclidean spaces  $\mathbb{E}^n$ for $n = 2$ and $3$ \cite{NWE90, FMSW65, MSVW67}. The classical or quantum Hamiltonian  (\ref{E3})
is said to be {\em superintegrable}
 if it admits $n+k$, $1 \le k \le n-1$ integrals of motion, $n$ of them in involution. It is {\em minimally superintegrable}
 for $ k=1$ and {\em maximally superintegrable} for $k = n-1$. For $n=2$ the two cases coincide and superintegrability
simply means the existence of three functionally independent integrals of motion (including the Hamiltonian).  For $n=3$ a superintegrable
system can have either 4 or 5 functionally independent integrals of motion. 

The $N$ body Calogero model \cite{FC71} (and, in particular, the three body one \cite{FC69}) 
is known to be superintegrable \cite{MA77, BCR00, HMS05, MFR99, Wa00, SW83}. An extensive literature 
exists on superintegrability in classical and quantum systems of the form (\ref{E3}) (see \cite{KKM05,RW02,TW04} and 
the references therein) devoted mainly, though not exclusively \cite{GW02,SG04} to systems
 with integrals  of motion of at most second order in the momenta. Superintegrable systems with 
complete sets of commuting quadratic integrals of motion are {\em multiseparable}. This means
that the corresponding Hamilton-Jacobi, or Schr\"{o}dinger equation allows the separation of 
variables in more than one system of
(orthogonal) coordinates. Alternatively, multiseparability can be described in terms of the geometric properties of the Killing 
two-tensors determined by the first integrals of motion that are quadratic in the momenta (see \cite{HMS05} as well as the relevant
references therein).

In what follows, we shall deal with the quantum mechanical problem, but all conclusions are the same (mutatis mutandis) for
the classical ones. For the systems admitting integrals of motion of order three or higher, this is not necessarily the case 
\cite{GW02,SG04, JH89}. 

\section{The Calogero model in the classification of superintegrable systems} 

In a recent article  \cite{HMS05} the {\em invariant theory of Killing tensors} (see also \cite{JMP02, JMP04, WF65} 
and the relevant references therein) was used to classify orthogonally separable Hamiltonian systems in the Euclidean
space $\mathbb{E}^3$. In particular, it was shown that the inverse square Calogero model with the potential 
\begin{equation}
\label{Calogero}
V = \frac{1}{(x_1-x_2)^2} + \frac{1}{(x_2-x_3)^2} + \frac{1}{(x_3-x_1)^2}
\end{equation}
allows the (orthogonal) separation of variables in 5 different coordinate systems, namely spherical, circular cylindrical, rotational parabolic, prolate
spheroidal and oblate spheroidal (see also \cite{BCR00,Wa00}). 

In this study \cite{HMS05} the potential (\ref{Calogero}) was viewed as a potential in the Hamiltonian (\ref{E3}), corresponding
to  a single particle in a potential field in $\mathbb{E}^3$. The potential (\ref{Calogero}) was shown  to allow 5 functionally
independent first integrals (including the Hamiltonian). From them it is possible to construct 5 inequivalent 
pairs of integrals in involution (in addition to the Hamiltonian). Each such pair is determined by two Killing tensors that
share the same orthogonal eigenvectors, thus generating an orthogonal separable system of coordinates. For example, 
the spherical coordinate
system is generated by the following pencil of Killing tensors (including the metric) whose components 
 given in terms of the Cartesian 
coordinates $(x_1,x_2,x_3)$ are as follows \cite{HMS05}: 
\begin{equation}
    \left[ \begin{array}{ccc} a_{1} + c_{2} x_3^{2} + c_{3} x_2^{2} & -c_{3} x_1x_2
      & -c_{2} x_1x_3 \\ [0.3cm]
-c_{3} x_1x_2 & a_{1} + c_{3} x_1^{2} + c_{2} x_3^{2} & -c_{2}
      x_2x_3 \\ [0.3cm]
-c_{2} x_1x_3 & -c_{2} x_2x_3 & a_{1} + c_{2} x_1^{2} + c_{2} x_2^{2}
      \end{array}\right]. \label{spherical}
  \end{equation}
The formula (\ref{spherical}) can be rewritten as  
\begin{equation}
\label{KTs}
a_1{g}^{ij} + c_2{K}^{ij}_1 + c_3{K}^{ij}_2, \quad i,j = 1,2,3, 
\end{equation}
 where  ${K}^{ij}_1$ and ${K}^{ij}_2$ are the components of  two {\em canonical} Killing tensors $\vec{K}_1$, $\vec{K}_2$ that share the same orthogonally integrable (i.e., surface
 forming) eigenvectors and ${g}^{ij}$ are the components of  the metric $\vec{g}$ of 
$\mathbb{E}^3$ (see \cite{HMS05} for more details). 

That notwithstanding, the Calogero potential (\ref{Calogero}) does not appear (at least explicitly)
 in the list of superintegrable
systems  in $\mathbb{E}^3$, established earlier \cite{NWE90,MSVW67}
 under the assumption that the first integrals that afford maximal or minimal 
superintegrability were to be quadratic in the momenta.  To unravel this mystery we first observe that the Killing tensors that determine the corresponding integrals of motion 
obtained for the
potential (\ref{Calogero})  in \cite{HMS05} are not in a {\em canonical} form (as in (\ref{spherical}), for example),
 but are rotated with respect to this form. As an example, 
let us consider again spherical coordinates $(r,\theta, \phi)$ in $\mathbb{E}^3$ generated by the hypersurfaces of the orthogonally integrable
eigenvectors of the Killing tensor (\ref{spherical}) given by the following coordinate transformations to the Cartesian 
coordinates $(x_1,x_2,x_3)$:  
\begin{equation}
x_1  = r\sin\theta\cos\phi, \quad x_2 = r\sin\theta\sin\phi, \quad x_3 = r\cos\theta.
\label{SC}
\end{equation}
A potential that allows separation in these coordinates must have the form 
\begin{equation}
V(r,\theta, \phi) = f(r) + \frac{1}{r^2}g(\theta) + \frac{1}{r^2\sin^2\theta} k(\phi)
\label{PSC}
\end{equation}
and the corresponding additional integrals of motion quadratic in the momenta will be in their standard form, namely
\begin{equation}
\begin{array}{rcl}
F_1 & = & \displaystyle  L_1^2 + L_2^2 + L^2_3 + 2\left[g(\theta) + \frac{1}{\sin^2\theta}k(\phi)\right], \\[0.3cm]
F_2 & = & L^2_3 + 2k(\phi),
\end{array}
\label{FI}
\end{equation}
where $L_i, i=1,2,3$ are the infinitesimal generators of $SO(3)$, that can be determined in terms of the Cartesian coordinates $x_i,$ $i = 1,2,3$ as follows: 
$L_1 = x_2p_3 - x_3p_2,$ $L_2 = x_3p_1 - x_1p_3,$ $L_3 = x_1p_2 - x_2p_1$. 
Note that the first integrals (\ref{FI}) in terms of the Cartesian coordinates  can be rewritten as 
\begin{equation}
\begin{array}{rcl}
F_1 & = &   K_1^{ij}p_ip_j + U_1(x_1,x_2,x_3), \\[0.3cm]
F_2 & = &   K_2^{ij}p_ip_j + U_2(x_1,x_2,x_3),
\end{array}
\label{FII}
\end{equation}
where $i,j = 1,2,3$,  $K^{ij}_1$, $K^{ij}_2$ are the components of the ``spherical'' Killing tensors (\ref{KTs}) 
and $(p_1,p_2,p_3)$ 
are the operators $(\frac{\partial}{\partial x_1}, \frac{\partial}{\partial x_2}, \frac{\partial}{\partial x_3})$
 respectively (quantum mechanics case) or the momenta components corresponding to the Cartesian coordinates 
$(x_1,x_2,x_3)$ (classical mechanics case). 

 If we rotate the $x_1-$, $x_2-$ and $x_3-$axes in (\ref{SC}), 
the form of the potential (\ref{PSC}) changes, so do the integrals (\ref{FI}), but separation of variables will 
still occur (in spherical coordinates with different axes). 

In the case of the potential (\ref{Calogero}) the rotation taking the Killing tensors into their standard form is a
non-trivial one, given by \cite{HMS05} (compare with (\ref{JC})): 

\begin{equation}
  \begin{pmatrix} x_1 \\x_2\\x_3 \end{pmatrix} = \frac{1}{\sqrt{6}} \begin{pmatrix} 2 & 0 & \sqrt{2} \\
    -1 & \sqrt{3} & \sqrt{2} \\ -1 & -\sqrt{3} & \sqrt{2} \end{pmatrix} \begin{pmatrix}\tilde{x}_1\\ \tilde{x}_2 \\
\tilde{x}_3 \end{pmatrix}.
     \label{TR}
\end{equation}
Accordingly, for the Calogero potential (\ref{Calogero}) we obtain
\begin{equation}
\label{Calogero1} 
V = 2\left[\frac{1}{(\sqrt{3} \tilde{x}_1 - \tilde{x}_2)^2}  +  \frac{1}{(\sqrt{3} \tilde{x}_1 + \tilde{x}_2)^2}
+ \frac{1}{\tilde{x}_2^2} \right]
\end{equation} 
and we see that the variable $\tilde{x}_3$ is absent from (\ref{Calogero1}). 
Expressing $\tilde{x}_1$ and $\tilde{x}_2$ in terms of spherical coordinates (\ref{SC}), we get
\begin{equation}
V = \displaystyle \frac{2}{r^2\sin^2\theta}\left[\frac{1}{(\sqrt{3}\cos\phi - \sin\phi)^2}
+ \frac{1}{(\sqrt{3}\cos\phi + \sin\phi)^2}
+ \frac{1}{\sin^2\phi} \right],
\label{CSP}
\end{equation}
i.e. a potential in the form (\ref{PSC}) with $f(r) = 0$, $g(\theta)=0$ and $k(\phi)$ specified. 

In what follows we show that after the rotation (\ref{TR}) it is possible to see that the Calogero potential (\ref{Calogero1}) is a member of an infinite
family of potentials, depending on one arbitrary function and sharing a number of important properties, such as superintegrability. Indeed, recall that all superintegrable potentials that
separate in spherical coordinates plus at least one other system were derived in \cite{MSVW67}. The potential 
\begin{equation}
\label{SPP}
V = \frac{k(\phi)}{r^2\sin^2\theta} 
\end{equation}
occurs several times. In what follows we list 5 {\em functionally independent}  first integrals (including the Hamiltonian $H$) that afford multi-separability for 
the potential (\ref{SPP}): 
\begin{equation}
\begin{array}{rcl}
H & = & \displaystyle \frac{1}{2}(p_1^2 + p_2^2 + p_3^2) + \frac{k(\phi)}{r^2\sin^2\theta}, \\[0.3cm]
F_1 & = & \displaystyle  L_1^2 + L_2^2 + L_3^2 + \frac{2k(\phi)}{\sin^2\theta}, \\[0.3cm]
F_2 & = & \displaystyle L_3^2  + 2k(\phi), \\[0.3cm] 
F_3 & = & \displaystyle \frac{1}{2} p_3^2, \\[0.3cm] 
F_4 & = & \displaystyle L_1p_2 + p_2L_1 - p_1L_2 - L_2p_1 -4 \frac{\cos\theta}{r\sin^2\theta} k(\phi),
\end{array}
\label{FIS}
\end{equation} 
where $k(\phi)$ is an arbitrary function. The functional independence of the first integrals (\ref{FIS}) has been verified with the aid of a computer algebra package (i.e.,  the 
Jacobian $\frac{\partial(H,F_1,F_2, F_3,F_4)}{\partial (x_1,x_2,x_3,p_1,p_2,p_3)}$ is of rank 5
at a generic point). 
It is important to note that the functionally independent first integrals (\ref{FIS}) are {\em linearly connected}, which means that they are subject to an additional constraint specified by the following expression in terms of the coordinates $\vec{x} = (x_1,x_2,x_3)$:
\begin{equation}
\label{const}
f_0(\vec{x}) H + f_1(\vec{x})F_1 + f_2(\vec{x})F_2 + f_3(\vec{x})F_3 + f_4(\vec{x})F_4 = 0,
\end{equation}
where $f_0(\vec{x})= 2x_3^2$, $f_1 (\vec{x}) = 1,$ $f_2(\vec{x}) = -1$, $f_3 (\vec{x}) =
-2(x_1^2 + x_2^2 + x_3^2)$, $f_4 = x_3.$ 
This formula is a consequence of the following ``rotational'' symmetry, that can be defined in a coordinate-free way. We can write all of the expressions in the formula (\ref{FIS}) as
$F_k = K_k^{ij}p_ip_j + U_k,$ where  $i,j = 1,2,3$. Then the Killing tensor $\vec{K}_k$ with the components $K^{ij}_k$ (including  the metric)  is   subject to the following formula
\begin{equation}
\label{RSC}
{\cal L}_{L_3}\vec{K}_k = 0,
\end{equation}
where ${\cal L}$ denotes the Lie derivative. We also note that the vector space spanned by the quadratic parts of the first integrals (\ref{FIS}) are invariant with  respect to translations along the $x_3-$axis. 
    
      It is easy to show now that the potential (\ref{SPP})
is orthogonally separable with respect to other systems of coordinates as well. Indeed, the pairs of involutive first integrals leading to the orthogonal separation 
of variables in the Schr\"{o}dinger equation are $\{F_1,F_2\}$ (spherical), $\{F_2, F_3\}$ (circular cylindrical), 
$\{F_2, F_4\}$ (rotational parabolic), and $\{F_2, F_1 \mp a^2 2F_3\}$ (oblate and prolate spheroidal). Another way to see
this is by looking at the separable potentials derived in \cite{MSVW67}. 
In terms of  Cartesian coordinates the potential (\ref{SPP}) is given by: 
\begin{equation}
V = \frac{k(x_2/x_1)}{x_1^2 + x_2^2}.
\label{CCP}
\end{equation}
Recall \cite{MSVW67} that the separable potentials corresponding
to ``rotational'' coordinates, namely spherical, circular cylindrical, rotational parabolic, oblate and prolate spheroidal in 
the Cartesian coordinates $(x_1,x_2,x_3)$ all have the
form 
\begin{equation}
V = f + g  + \frac{k(x_2/x_1)}{x_1^2 + x_2^2}, 
\label{GP}
\end{equation}
where  $k$ are arbitrary functions, while $f$ and $g$ are specified differently in each case. The
 common part of the five separable potentials is exactly the potential (\ref{CCP}).

These observations put in
evidence that the potential (\ref{CCP}) defines a family of  maximally
 superintegrable potentials separable with respect
to the five ``rotational'' orthogonal coordinate systems, namely spherical, circular cylindrical, rotational parabolic, 
oblate and prolate spheroidal whose Killing tensors are constrained by the rotational symmetry condition (\ref{RSC}). As for the Calogero potential (\ref{Calogero1}),
 in the coordinates $(\tilde{x}_1,\tilde{x}_2,
\tilde{x}_3)$ determined by the transformation (\ref{CSP}), it assumes  the form (\ref{CCP}) for 
\begin{equation}
\label{k}
k(t) = \displaystyle 2(1+t^2)\left[\frac{3 + t^2}{(3 - t^2)^2} + 1\right], 
\end{equation}
where $t = \tilde{x}_2/\tilde{x}_1$. 

The potential (\ref{SPP}) can be imbedded into more general families of potentials in $\mathbb{E}^3$ that are 
{\em minimally} superintegrable. In contrast to maximally superintegrable potentials
they admit three additional integrals rather than four. They are 
\begin{equation}
\begin{array}{rcl}
V_1 & = & \displaystyle  \alpha(x_1^2 + x_2^2 + x_3^2) + \frac{\beta}{x_3^2} + \frac{1}{x_1^2 +x_2^2}h(\phi), \\[0.3cm]
V_2 & = & \displaystyle  \frac{\alpha}{r} + \beta\frac{\cos\theta}{r^2\sin^2\theta} + \frac{1}{r^2\sin^2\theta}h(\phi), \\[0.3cm]
V_3 & = & \displaystyle  k(x_1^2 + x_2^2) + 4kx_3^2 + \frac{1}{x_1^2 +x_2^2}h(\phi). 
\end{array}
\end{equation}
The potential $V_1$ with $(\alpha, \beta) \not= (0,0)$ separates in all of the 5  ``rotational'' coordinate
systems considered above except rotational parabolic ones. $V_2$ separates only in spherical  and rotational
parabolic, while $V_3$ - in cylindrical  and rotational parabolic. We mention that a special case
of $V_2$ with $\beta = 0$ and $h(\phi) = \mbox{const}$ is the Hartmann potential used  in molecular 
physics to describe ring-shaped molecules \cite{HH72, KW87}. 

The rotation (\ref{TR}) in $\mathbb{E}^3$ has a simple  meaning for three particles on a line
with inverse square potentials. Comparing (\ref{E2}) with  (\ref{CSP}), we see that
the rotation corresponds to introducing centre-of-mass coordinates (\ref{JC}). If we factor out
the centre-of-mass motion (i.e. drop the term $\frac{1}{2}p_3^2$ in the kinetic energy), we reobtain
the Hamiltonian (\ref{E2}) with $\omega = 0$. 

The system  (\ref{E2}) can be viewed as one particle in a potential in the Euclidean plane $\mathbb{E}^2$. Interestingly, 
it is not multiseparable. For both  $\omega = 0$ and  $\omega \not= 0$ it separates only in polar
coordinates, so it allows only one second order integral of motion (in addition to the Hamiltonian), namely
\begin{equation}
\label{F}
F = \displaystyle  L_3^2 - \left[\frac{g_1}{(\sqrt{3}\sin\phi - \cos\phi)^2}
 + \frac{g_2}{(\sqrt{3}\sin\phi + \cos\phi)^2} + \frac{g_3}{\cos^2\phi} \right].
\end{equation}
If the system (\ref{E2}) is superintegrable in $\mathbb{E}^2$, the second integral of motion must be of
higher order in the momenta, not commuting with $F$ given by (\ref{F}). Multiseparability of  a physical system, in particular the Calogero 
model, may also be of interest from the point of view of different 
possible quantizations. In a recent article F\'{e}her {\em et al} \cite{FTF05} have used separation of variables in circular
cylindrical 
coordinates in the three-body Calogero model to investigate all possible 
self-adjoint extensions of the corresponding angular and radial 
Hamiltonians. The question arises whether separation of variables in other 
coordinates might not lead to different quantizations. 

\section{Conclusions}

The beauty of the Calogero model is lost when its potential is written in the form (\ref{Calogero1}). 
The formula (\ref{Calogero1})
does however show that  this system is a member of a family of maximally superintegrable systems determined 
by the general formula (\ref{SPP}), involving  an arbitrary function of one variable, the azimuthal angle
$\phi$. All of them allow the orthogonal  separation of variables in the 5 different ``rotational'' coordinate systems. 
The complete set of commuting operators (first integrals) in each case consists of the Hamiltonian $H$ and 
$F_2$ of (\ref{FIS}) and one more operator ($F_1, F_3, F_4$ and $F_1\mp a^2p_3^2$, respectively). The
operator $F_2$ that is thus singled out corresponds, in the case of the free motion, to a one-dimensional
subgroup of the (orientation-preserving) isometry group $I(\mathbb{E}^3)$, which is the symmetry group
 of the Schr\"{o}dinger equation without a potential. This subgroup generates the angle $\phi$, common to 
all 5 ``rotational'' orthogonally separable coordinate systems. 

This raises the question whether other maximally superintegrable systems involving arbitrary functions exist. 
All superintegrable systems in $\mathbb{E}^3$ separating in spherical coordinates and in one further 
system were found in \cite{MSVW67}. All further systems separable in (at least)
two coordinate systems were found in \cite{NWE90}. In the lists provided by Evans \cite{NWE90}
five systems are maximally superintegrable and each one depends on artibrary constants. 
In addition, eight systems are listed as minimally superintegrable, each depending on one
arbitrary function and up to three constants. One of the minimally superintegrable systems has the 
potential
\begin{equation}
\label{V1}
V_1 = F(r) + \frac{c_1}{x_1^2} + \frac{c_2}{x_2^2} + \frac{c_3}{x_3^2},
\end{equation}
where $c_1,c_2$ and $c_3$ are arbitrary constants. Here and below $r,\theta$ and $\phi$ are
spherical coordinates as specified by (\ref{SC}). Its superintegrability is due to the fact 
that the corresponding Hamiltonian commutes with the operators 
\begin{equation}
\begin{array}{rcl}
F_1 & = & \displaystyle L_1^2 + \frac{2c_2\cos^2\theta}{\sin^2\theta\sin^2\phi} + \frac{2c_3\sin^2\theta \sin^2\phi}{\cos^2\theta}, \\[0.3cm]
F_2 & = & \displaystyle L_2^2 + \frac{2c_1\cos^2\theta}{\sin^2\theta \cos^2\phi} + \frac{2c_3\sin^2\theta \cos^2\phi}{\cos^2\theta}, \\[0.3cm]
F_3 & = & \displaystyle L_3^2 + \frac{2c_1}{\cos^2\phi} + \frac{2c_2}{\sin^2\phi}.
\end{array}
\end{equation}
This potential becomes maximally superintegrable for $F = \omega(x_1^2 + x_2^2 + x_3^2)$. For $c_1 = c_2 = c_3 = 0$
it simply becomes rotationally invariant (but not maximally superintegrable). Four of the minimally superintegrable
potentials have the form 
\begin{equation}
\label{Vi}
V_i(x_1,x_2,x_3) = \tilde{V}_i(x_1,x_2) + f(x_3), \quad i = 2,3,4,5, 
\end{equation}
where $\tilde{V}_i(x,y)$ is one of the four multiseparable potentials in $\mathbb{E}^2$ \cite{FMSW65}. In each
case the set of integrals of motion consists of
\begin{equation}
\label{F1}
F_1 = \frac{1}{2}p_3^2 + f(x_3)
\end{equation}
and three  further operators, the principal parts of which lie in the enveloping algebra of the Lie algebra of the isometry group
$I(\mathbb{E}^2)$. In particular, for $\tilde{V}_i(x_1,x_2) = 0$ the Hamiltonian  and $F_1$ of (\ref{F1})
commutes with the Lie algebra $\{L_3, p_1, p_2\}$, i.e. $H$ and $F_1$ are invariant under the orientation-preserving
isometry group $I(\mathbb{E}^2)$. This provides a total of 4 integrals of motion, never 5. 
Out of these 4 functionally independent integrals of motion we can form 
4 inequivalent triplets of integrals of motion in involution, namely $(H, 
F_1, X_i),$ $i=1,2,3,4$ with
$$X_1=p_1^2, \quad X_2=L_3^2,\quad X_3= L_3 p_1 +p_1 L_3,\quad X_4 = L_3^2 +a^2( p^2_1 - p^2_2),$$ 
where $a\not=0$.These triplets correspond to the separation of variables in the cartesian, 
polar, parabolic translational and elliptic translational coordinates,
respectively. Within the $x_1x_2-$plane the origin and the orientation of axes 
can be chosen arbitrarily.

Finally, three of the minimally superintegrable systems depend on an arbitrary function of the azimuthal
angle $\phi$. They all have the form
\begin{equation}
\label{Ve}
V_i (r,\theta, \phi) = \tilde{V}_i(r,\theta) + \frac{k(\phi)}{r^2\sin^2\theta}, \quad i = 4,7, 8. 
\end{equation}
The integral $F_2$ of (\ref{FIS}) is present in each case, together with $H$ and one of 
$F_1, F_3$ or $F_4$. In particular,  for $\tilde{V}_i(r, \theta) = 0$ all of the operators (\ref{FIS})
are integrals of motion. 

We conclude that in $\mathbb{E}^3$ the potential (\ref{SPP}) is the only potential that
is maximally superintegrable and depends on an arbitrary function (of one variable). The three 
body Calogero model corresponds to one particular choice of this function, namely that given in (\ref{SPP}) and (\ref{k}). 

An important question arises in this context. Namely, what are the physical consequences in classical and quantum mechanics, of the
existence of a maximally superintegrable system, depending on an arbitrary function? In classical mechanics maximally superintegrable
sysetms have the property that their finite trajectories are closed  \cite{Ne72}. In quantum mechanics
they have degenerate energy levels and it has been conjectured \cite{TTW01,  RW02} that they are
exactly solvable. We cannot expect these properties to hold for the potential (\ref{SPP}) with $k(\phi)$ arbitrary.  
We suspect that the reason for this paradox is that the 5 integrals (\ref{FIS}) are functionally independent, but linearly connected. 

One of the messages that we arrive at is that results considered to be ``canonical'' in one approach
to a problem may be quite non-obvious in another. Thus, the Killing tensors obtained in \cite{HMS05} were
not in canonical (standard) form for the Calogero model viewed as an $\mathbb{E}^3$ problem. The
advantage of the invariant approach used in \cite{HMS05, JMP02, JMP04, WF65} is the following.  For a given isometry group action in
a vector space of Killing tensors  one can employ the approach developed in \cite{HMS05, JMP02, JMP04, WF65} to
determine which orbit a  Killing tensor belongs to and then find the corresponding isometry group action mapping the Killing tensor in
question
to  its canonical form (i.e., the corresponding {\em moving frames map}).

\bigskip 

\noindent {\bf Acknowledgements.} The authors' research was partially supported by Discovery Grants from NSERC of 
Canada.

\end{document}